\documentclass[aps,showpacs,preprintnumbers]{revtex4}
\usepackage{bm}

\input{tcilatex}

\begin{document}

\title{Angular momentum quantization from Planck 's energy quantization}
\author{J.H.O.Sales}
\affiliation{$^{1}$Funda\c{c}\~{a}o de Ensino e Pesquisa de Itajub\'{a} , Av. Dr. Antonio
Braga Filho, CEP 37501-002, Itajub\'{a}, MG.\\
$^{2}$Instituto de Ci\^{e}ncias, Universidade Federal de Itajub\'{a}, CEP
37500-000, Itajub\'{a}, MG, Brazil.}
\author{A.T. Suzuki}
\affiliation{Instituto de F\'{\i}sica Te\'{o}rica-UNESP, Rua Pamplona 145, CEP\ 01405-900
S\~{a}o Paulo, SP, Brazil}
\author{D.S. Bonaf\'{e}}
\affiliation{Escola Estadual Major Jo\~{a}o Pereira, Av. Paulo Chiaradia, Cep: 37500-028,
Itajub\'{a}, MG, Brazil.}
\date{\today }

\begin{abstract}
We present in this work a pedagogical way of quantizing the atomic orbit for
the hydrogen's atom model proposed by Bohr without using his hypothesis of
angular momentum quantization. In contrast to the usual treatment for the
orbital quantization, we show that using energy conservation, correspondence
principle and Plank 's energy quantization Bohr's hypothesis can be deduced
from and is a consequence of the Planck's energy quantization.
\end{abstract}

\maketitle







\section{Introduction}

\noindent The existence of an atomic nucleus was confirmed in 1911 by E.
Rutherford in his classic scattering experiment \cite{ruth}. Before that
time, it was believed that an atom was something like a positively charged
dough and negatively charged raisins scattered here and there within the
dough, or an ice-cream with chocolate chip flakes in it, where the ice-cream
would be the positive charge (protons) and the chocolate chip flakes the
negative charge (electrons); an atomic model proposed by J.J.Thomson in
1904. However, this atomic model had many problems. And it became untenable
as E. Rutherford showed the inconsistencies of such a model in face of his
experiments of alpha particle (ionized helium) scattering by thin sheets of
gold as targets. The main objection to that model being that scattering
experiments indicated a far more ''dilute'' type of matter constituints and
''empty'' space within objects.

Then, Rutherford himself proposed the orbital model for the atom, in which
there was a central nucleus populated by positively charged particles --
protons -- and negatively charge particles -- electrons -- moving around the
nucleus in orbital trajectories, similar to the solar system with the sun at
the center and planets orbiting it, described by the mechanics of celestial
bodies acted on under central forces among them. Later on, the concept of
neutral particles -- neutrons -- within the nucleus came to be added into
the model.

The planetary model, however, was not free of problems. The main objection
was that in such a model, electrons moving around the nucleus would radiate
energy and therefore, classically, such an atom would collapse into itself
after electrons radiated all their energy. Therefore, the model proposed by
Rutherford had to be modified.

This modification came through the introduction of a completely radical
concept unknown to classical physics, i.e., the quantization of physical
quantities. In 1912 N. Bjerrum and a year later P.\ Ehrenfest were the first
ones to apply successfully the quantum principles to the study of molecular
spectroscopy. Following the suggestion of Lorentz and inspiring in the ideas
of Nernst, they considered that a diatomic molecule could be seen as two
atoms oscillating along the line that links the two, and that this line also
rotates in a plan. This last rotational movement they supposed to be such
that its energy would be multiple of $hf$ \ (according to Bjerrum) or $\frac{%
1}{2}hf$ (according to a later correction by Ehrenfest), $f$ being the
number of revolutions per second the molecule executed in this plan of
rotation.

Two major changes in the building and understanding of the atomic models
were developed at the beginning of XX century: firstly it was necessary to
forgo the idea that the lines of the atomic spectra were due to the natural
modes of atomic oscillations, and secondly that the spectral lines had to be
seen as quantum phenomena or quantum effects.

In this line, Bohr was the first one who was able to explain the Balmer
series of the atomic spectra by postulating the angular momentum
quantization for the hydrogen atom 's orbitals.

We want to emphasize that in this hystoric development of the atomic model
the quantization of the angular momentum was necessary to ensure the
stability of the atomic orbital. In our work here, we start of imposing the
stability condition for the atomic orbital and demonstrate that requiring
the stable minimum for the Planck 's energy quantization, we arrive at the
angular momentum quantization.

In the next section we present the Rutherford's orbital model as a general
introduction. In section 3 we present Bohr's hypothesis for quantizing
atomic orbitals. Section 4 is devoted to developing our main ideas
concerning the requirement to minimize the Planck's energy and use it to
identify with the Bohr's quantization hypothesis. Finally the last section
is for our conclusions.

\section{The planetary atomic model}

\noindent The discoveries that ocurred by the end of XIX century led the
physicist Ernest Rutherford to do scattering experiments that culminated in
a proposal for the planetary model for atoms.

According to this model, all positive charge of a given atom, with
approximately $99\% $ of its mass, would be concentrated in the atomic
nucleus. Electrons would be moving around the nucleus in circular orbits and
these would be the carriers of the negative charges. Knowing that the charge
of an electron and the charge of a proton are the same in modulus, and that
the nucleus has $Z$ protons, we can define the charge of the nucleus as $%
Ze^{-}$.

Experimentally we observe that in an atom the distance $r$ between the
electron orbit and the nucleus is of the order $10^{-10}m$.

In this section, we build the planetary model for the atom and analyse its
predictions compared to experimental data.

Using Coulomb's law 
\begin{equation}
F=\frac{Z\left\vert e^{-}\right\vert ^{2}}{4\pi \varepsilon _{0}r^{2}}
\label{1}
\end{equation}
and the centripetal force acting on the electron in its circular orbit 
\begin{equation}
F_{c}=\frac{m_{e}v^{2}}{r}  \label{2}
\end{equation}%
results in 
\begin{equation}
\frac{Z\left\vert e^{-}\right\vert ^{2}}{4\pi \varepsilon _{0}r^{2}}=\frac{%
m_{e}.v^{2}}{r}  \label{3}
\end{equation}%
\begin{equation}
v=\sqrt{\frac{e^{2}}{4\pi \varepsilon _{0}m_{e}r}.Z}  \label{4}
\end{equation}%
where we have used the shorthand notation $\left\vert e^{-}\right\vert =e$.

From (\ref{4}) we can estimate the radius for the electron orbit, 
\begin{equation}
r=\frac{e^{2}}{4\pi \varepsilon _{0}m_{e}v^{2}}Z\text{ ,}  \label{4a}
\end{equation}
which means that the radius depends on the total number of protons in the
nucleus, $Z$, and also on the electron's velocity. Here we can make some
definite estimates and see whether our estimates are reasonable, i.e.,
agrees or does not violate experimental data. According to the special
theory of relativity, no greater velocity can any particle possess than the
speed of light. More precisely, for particles with mass like electrons, we
know that their velocity is limited by $v < c$ where $c$ is the speed of
light in vacuum. Substituting for velocity $v=c=3\times10^8 $m/s, electric
charge $e=1.602\times 10^{-19}$m/s, mass of the electron $m_{e}=9.109\times
10^{-31}$kg and $\varepsilon _{0}=8.854\times 10^{-12}$F/m \cite{1a}, we
have 
\[
r>2.813\times 10^{-15}\text{m,} 
\]
where we have taken $Z=1$ and didn't consider the relativistic mass. This
means that we have a lower limit for the radius of an electron's orbit
around the central nucleus, which is consistent with the experimental
observed data where radius of electronic orbits are typically of the order $%
10^{-10}$m.

\subsection{Limitations of the model}

Even though this model could explain some features of the atomic structure
concerning the scattering data, there was nonetheless problems that could
not be explained just by classical mechanics analysis. Since protons and
electrons are charged particles, electromagnetic forces do play their role
in this interaction, and according to Maxwell's equations, an accelerated
electron emits radiation (and therefore energy), so that electrons moving
around the nucleus would be emitting energy. This radiated energy would of
course lead to the downspiraling of electrons around the nucleus until
hitting it. Classical theoretical calculations done predicted that all
electrons orbiting around a nucleus would hit it in less than a second!

However, what we observe is that there is electronic stability, and
therefore the model had to be reviewed.

\section{Bohr's hypothesis}

Analysis of the hydrogen spectrum which showed that only light at certain
definite frequencies and energies were emitted led Niels Bohr to postulate
that the circular orbit of the electron around the nucleus is quantized,
that is, that its angular momentum could only have certain discrete values,
these being integer multiples of a certain basic value \cite{04}. This was
his \emph{\textquotedblleft ad hoc\textquotedblright } assumption,
introduced by hand into the theory. In 1913, therefore, he proposed the
following for the atomic model \cite{Bohr}:

\noindent 1. The atom would be composed of a central nucleus where the
positive charges (protons) are located;

\noindent 2. Around the central nucleus revolved the electrons in equal
number as the positive charges present in the nucleus. The electrons
orbiting such a nucleus had discrete quantized energies, which meant that
not any orbit is allowed but only certain specific ones satisfying the
energy quantization requeriments;

\noindent 3. The allowed orbits also would have quantized or discrete values
for orbital angular momentum, according to the prescription $|\mathbf{L}%
|=n\hbar $ where $\hbar =\displaystyle\frac{h}{2\pi }$ and $n=1,2,3,...$,
which meant the electron's orbit would have specific minimum radius,
corresponding to the angular momentum quantum number $n=1$. That would solve
the problem of collapsing electrons into the nucleus.

Two colloraries for Bohr's assumptions do follow: First, from item 2. above,
the laws of classical mechanics cannot describe the transition of an
electron from one orbit to another, and second, when electrons do make a
transition from one orbit to another, the energy difference is either
supplied (transition from lower to higher energy orbits) or carried away
(transition from higher to lower energy orbits) in discrete values. Only
decades later (1926) G.N.Lewis coined the name \textit{photon }for the
particle carrier of this quantum of energy\textit{.}

First, let us follow the usual pathway where Bohr's quantization is
introduced. Using Newton's second law for the electron moving in a circular
orbit around the nucleus, and thus subjec to Coulomb's law, we have: 
\begin{equation}
\frac{e^{2}}{4\pi \varepsilon _{0}r^{2}}=m\frac{v^{2}}{r}.  \label{c1}
\end{equation}

This allows us to calculate the kinetic energy of the electron in such an
orbit: 
\begin{equation}
E_{\text{k}}=\frac{1}{2}mv^{2}=\frac{e^{2}}{8\pi \varepsilon _{0}r}.
\label{ec1}
\end{equation}

The potential energy for the system proton-electron on the other hand is
given by 
\begin{equation}
E_{\text{p}}=-\frac{e^{2}}{4\pi \varepsilon _{0}r},  \label{ep}
\end{equation}
where $r$ is the radius of the electronic orbit.

Therefore, the total energy for the system is 
\begin{equation}
E=-\frac{e^{2}}{8\pi \varepsilon _{0}r}.  \label{et}
\end{equation}

This result would suggest that, since the radius can have any value, the
same should happen with the angular momentum $L$. 
\begin{equation}
L=pr\sin \theta =pr,\text{ where }\theta =90^{0}  \label{l}
\end{equation}%
that is, the angular momentum depends on the radius. The linear momentum of
the electron is given by 
\begin{equation}
p=mv\text{.}  \label{p}
\end{equation}

Therefore the problem of quantizing the angular momentum $L$ reduces to the
quantizing of the radius $r$, which depends on the total energy (\ref{et}).
Just here Bohr introduced an \textit{additional hypothesis}, in that the
angular momentum of the electron is quantized, i.e., 
\begin{equation}
L_{\text{Bohr}}=n\hbar ,  \label{quantl}
\end{equation}%
where $\hbar =\displaystyle\frac{h}{2\pi }$. In this manner he was able to
quantize the other physical quantities such as the total energy. This is the
usual pathway wherein the textbooks normally follow in their sequence of
calculations.

\section{Origin of orbital quantization}

The total energy of the atom can be viewed from the dynamics of the atomic
model constituints, i.e., electron-proton (see section 2). According to the
deduction given in the Appendix, \textit{modulus its sign,} the total energy
for the system electron-proton is

\begin{equation}
E=\frac{mv^{2}}{2}.  \label{syst}
\end{equation}%
Note that we can accommodate either the plus or minus sign here, depending
on where we put the ground zero potential energy reference for the electron.
Knowing that the scalar orbital velocity of the electron is 
\begin{equation}
v=2\pi rf,  \label{r2a}
\end{equation}%
where $f$ is the orbital frequency and $r$ is the radius of the electronic
orbit. Substituting this $v$ into ( \ref{syst} ) we then have:

\begin{equation}
E=2\pi ^{2}mr^{2}f^{2}  \label{r2b}
\end{equation}%
so the ratio of the system's total energy variation with respect to the
frequency is 
\begin{equation}
\left. \frac{dE}{df}\right\vert _{\text{System}}=4\pi ^{2}mr^{2}f=2\pi (2\pi
rf)(mr)=2\pi mvr=2\pi pr=2\pi L  \label{r3}
\end{equation}

Here the variation of total energy with frequency is subscripted by "System"
because it comes from the classical dynamics of the system. Note too, that
this variation defines the minimum for the total energy since the second
derivative with respect to frequency is positive. Moreover, such variation
can be rewritten in such a way as to be proportional to the angular momentum 
$L$.

Now, following Planck, let us consider that in the interaction of radiation
(light) with matter, radiation 's energy is quantized according to

\begin{equation}
E=nh\nu  \label{planck}
\end{equation}%
where $n$ is a natural number ($n=1,2,3...$),\ $h$ is the Planck 's constant
and $\nu $\ \ is the frequency of the interacting radiation. As the
interaction occurs, this energy varies with respect to its frequency as: \ 
\begin{equation}
\left. \frac{dE}{d\nu }\right\vert _{\text{Planck}}=nh.  \label{deri}
\end{equation}

Here we note that in (\ref{r3}) $f$ is the frequency of the orbital movement
of the electron, while in (\ref{deri}) $\nu $ is the frequency of the
radiation, and they are, of course, different in principle, so we cannot
equate the two equations. It is here that Bohr 's \ "correspondence
principle" comes into play, according to which it is hypothesized that both
frequencies are the same, that is, $\nu =f$.

Using Bohr 's correspondence principle we can therefore write

\bigskip 
\begin{equation}
\left. \frac{dE}{df}\right\vert _{\text{Planck}}=nh.  \label{deri2}
\end{equation}

By energy conservation and correspondence principle it follows that 
\[
\left. \frac{dE}{df}\right\vert _{\text{System}}=\left. \frac{dE}{df}%
\right\vert _{\text{Planck}} 
\]%
from which

\begin{equation}
L=\frac{nh}{2\pi }.  \label{r5}
\end{equation}

This is exactly Bohr 's angular momentum quantization.

\section{Conclusions}

In our work here we have deduced the angular momentum quantization for
atomic orbitals firstly proposed and hypothesized by Bohr by making use of
three principles: energy conservation, correspondence principle and Planck's
quantization for the radiation interacting with matter.

Our deduction here therefore differs from the standard one where the
quantization for the angular momentum for atomic orbitals is arrived at by
comparing the energy of rotating electron with half of the energy of an
oscillator associated with it.

Instead of comparing the energies involved in the rotating electron with the
quantized energy for the oscillator associated with it, our starting point
is the energy conservation in the variation of energy of electron 's
movement with the variation of energy carried by the radiation interacting
with the atom, which is quantized according to Planck. This leads to the
angular momentum quantization.

In this work we have shown that Planck's fundamental assumption of energy
quantization is more fundamental than Bohr's assumption of angular momentum
quantization. In fact, we have shown that Bohr's rule for angular momentum
quantization can be derived from either assuming quantization for the
oscillator energy (traditional view) or by assuming that the variation of
energies as the atom interacts with radiation is quantized.

\section{Appendix}

In this Appendix we will show that the total energy for the system
electron-proton depends on the scalar orbital velocity of the elecron. From
classical mechanics, an electron orbiting a proton in a circular orbit obeys
the folowing equilibrium of forces:

\begin{equation}
\frac{e^{2}}{4\pi \varepsilon r^{2}}=\frac{mv^{2}}{r}  \label{a1}
\end{equation}%
which results in%
\begin{equation}
\frac{e^{2}}{4\pi \varepsilon r}=mv^{2}  \label{a2}
\end{equation}

The total energy $E$ for the electron-proton system is equal to the sum of
its kinetic energy $E_{k}$ and its potential energy $E_{p}$.

\begin{equation}
E=E_{\text{k}}+E_{\text{p}}  \label{a5}
\end{equation}

where

\begin{equation}
E_{\text{k}}=\frac{mv^{2}}{2}  \label{a6}
\end{equation}

and

\begin{equation}
E_{\text{p}}=-\frac{e^{2}}{4\pi \varepsilon r}  \label{a7}
\end{equation}

The negative sign for the potential energy indicates that our zero reference
for it is at infinity.

Substituting these in (\ref{a5}) we have:

\begin{equation}
E=\frac{mv^{2}}{2}-\frac{e^{2}}{4\pi \varepsilon r}  \label{a8}
\end{equation}

From (\ref{a2}), we obtain:

\[
E=\frac{mv^{2}}{2}-mv^{2}=-\frac{mv^{2}}{2} 
\]

Therefore, the total energy $E$ for the electron-proton system is:

\[
E=-\frac{mv^{2}}{2} 
\]

Again, the negative sign here defines the bound state of the atom.

\textbf{Acknowledgments:} D. S. Bonaf\'{e} thanks the PIBIC-Jr to the
Universidade Federal de Itajub\'{a}-MG and J.H.O. Sales from FAPEMIG-CEX
1661/05 and FMC Equipamentos Eletrom\'{e}dicos/FINEP. JHO Sales, thanks the
hospitality of the Institute for Theoretical Physics, UNESP, where part of
this work has been performed..

\end{document}